\documentclass[preprint,showpacs,preprintnumbers,amsmath,amssymb,showkeys]{revtex4}

\usepackage{graphicx}% Include figure files
\usepackage{dcolumn}% Align table columns on decimal point
\usepackage{bm}% bold math

\begin{document}

%\preprint{APS/123-QED}

\title{Local Inertial Coordinate System and The Principle of Equivalence}% Force line breaks with \\

\author{Xin Zhang}
 \email{xzhang_cn@yahoo.com}
\affiliation{%
Department of Physics, Nanjing University, Nanjing 210093, P.R.China
}

\author{Bin Xi}
 \email{bean@itp.ac.cn}
\affiliation{%
Key Laboratory of Frontiers in Theoretical Physics, Institute of
Theoretical Physics, CAS, P.O. Box 2735, Beijing 100080, P.R.China }

\date{\today}% It is always \today, today,
             %  but any date may be explicitly specified

\begin{abstract}
In this paper, the local inertial coordinate system is calculated
through coordinate transformations from laboratory coordinate
system. We derived the same free falling equations as those in
General Relativity. However, the definitions of second and higher
order covariant derivatives are different. Our results are different
from the classic view of the principle of equivalence, and suggest
that all the laws of physics in gravitational field can be given by
 doing coordinate transformations from local inertial coordinate system to
 lab coordinate system.
\end{abstract}

% PACS, the Physics and Astronomy
                             % Classification Scheme.
%Use showkeys class option if keyword
                              %display desired
\maketitle

\section{Introduction}
Einstein's General Relativity are in good agreement with the solar
observation data, but it also meets great challenges in galactic and
larger scales. So either we have to accept the existence of dark
matter and dark energy, or we must go another way to modify the
theory of Gravitation\cite{mond}\cite{san}\cite{ben}\cite{soti}. We
here, motivated by a theoretical approach, reconsidered the
translation of equivalence principle, the foundation of General
Relativity. It is well-known that General Relativity is based on the
principle of equivalence, ``\textit{at every space-time point in an
arbitrary gravitational field it is possible to choose a `locally
inertial coordinate system' such that, within a sufficiently small
region of the point in question, the laws of nature take the same
form as in unaccelerated Cartesian coordinate systems in the absence
of gravitation}''\cite{gac}. Equivalence principle describes the
gravitational effects as coordinate transformations, but General
Relativity does not. Our ultimate aim is to establish a coordinate
theory of gravitation. As a modest step, we here give the LICS
(local inertial coordinate system) a full description, and the
further topic concerning the gravitational field equations will be
given in a forth coming paper\cite{zx}. In our opinion, all the
gravitational effects can be described by the LICS, and the
gravitational equations should also have great relations to the
LICS, because the gravitational equations should become trivial if
we transform the coordinates from lab coordinate system to LICS. So
it is of essential importance to find the coordinate transformations
between the LICS and the lab coordinate system.

In the theory of General Relativity, the LICS at a spacetime point
$\mathcal{P}_0$ is treated as the ``closest thing"\cite{grav} to
global Lorentz frame:
\begin{equation}\label{local}
\left\{ \begin{array}{cc}
        g_{\mu\nu}(\mathcal{P}_0)=\eta_{\mu\nu} &  \\
        g_{\mu\nu,\alpha}(\mathcal{P}_0)=0 &
        \end{array} \right.
\end{equation}
 $g_{\mu\nu,\alpha,\beta}(\mathcal{P}_0)$ is non-zero due to the spacetime
 curvature. In this paper we should like to point out that we
 find the LICS in one point of gravitation is the ``same thing" to
global Lorentz frame. Symbols denoting local coordinate
transformations instead of metric tensors are used during
calculations. All the laws of physics in gravitational field can be
given by doing a certain coordinate transformation, so the gravitational
effects are treated totally as some kind of coordinate
transformations. It will be demonstrated that the free falling
equations of a testing neutral particle are in coincidence with the
well known ones
\begin{equation}
\frac{dU^{\lambda}}{d\tau}+\left\{ \begin{array}{cc} \lambda & \\
 \mu & \nu    \end{array}   \right\} U^{\mu \nu}=0,
\end{equation}
in which $\left\{ \begin{array}{cc} \lambda & \\
 \mu & \nu    \end{array}   \right\}$ is the Christoffel symbol.
 And we will continue show a different structure of the second and higher order covariant
 derivatives through coordinate transformations. The form of the second order covariant derivatives is
given and higher orders can be calculated in a straightforward way.
We also find that the Riemann curvature tensor
$R_{\kappa\omega\mu\nu}$, as the description of spacetime curvature,
is not sufficient. It is only a rank 2 description, and a full
definition is given in this paper. Given a point particle p of
infinitesimal mass in gravitational field, in the LICS of p, the
gravitational field equation should be second order differential
equation. With a coordinate transformation to lab coordinate system,
we can find the contribution of particle p to the gravitational
field. We do not concentrate on this point here, which is rather a
complicated problem, and it will be further discussed in \cite{zx}.
In general, we would like to emphasize the importance of finding a
detailed description of LICS and hence a different structure of
covariant derivatives.

\section{the Free Falling Equation}

In developing our new interpretation we start with a coordinate
transformation

\begin{equation}\theta^{\alpha}_{\phantom{a}\mu}(x)\equiv
\frac{\partial \xi^\alpha}{\partial
x^\mu}(x)=\Lambda^{\alpha}_{\phantom{a}\beta}(x)X^{\beta}_{\phantom{a}\mu}(x).
\end{equation}

Here $\{\xi^\alpha\}$ are the local Lorentz coordinates and
$\{x^\mu\}$ are the coordinates used by distant observers.
$\Lambda^{\alpha}_{\phantom{a}\beta}(x)$ are the local Lorentz
transformations, and $X^{\beta}_{\phantom{a}\mu}(x)$ denotes the
remain part of the coordinate transformations. Similarly we can also
define a reverse transformation of the coordinates

\begin{equation} e^{\mu}_{\phantom{a}\alpha}(x)\equiv
\theta^{-1 \mu}_{\phantom{-1a}\alpha}(x)=\frac{\partial
x^\mu}{\partial \xi^\alpha}=X^{-1
\mu}_{\phantom{-1a}\beta}(x)\Lambda^{\phantom{a}\beta}_{\alpha}
\end{equation}

in which $X^{-1 \mu}_{\phantom{-1a}\beta}$ is the inverse matrix of
$X^{\beta}_{\phantom{a}\mu}$, so is the matrix
$\Lambda^{-1\beta}_{\phantom{-1a}\alpha}=\Lambda^{\phantom{a}\beta}_{\alpha}
=\eta_{\alpha\gamma}\Lambda^{\gamma}_{\phantom{a}\delta}\eta^{\delta\beta}$.
The $X$ (or $X^{-1}$) part of coordinate transformations makes
inertial coordinates into non-inertial ones, so gravitational
redshifts are embed in the first introduced terms
$X^{\beta}_{\phantom{a}\mu}$ and $X^{-1 \mu}_{\phantom{-1a}\beta}$,
which also include arbitrary coordinate transformations of the
observers. In the gravitational field, the LICS can only exist
within a sufficiently small region of any point, as a result, a free
falling inertial coordinate system in the gravitational field must
have been ``twisted". So the $X^{\beta}_{\phantom{a}\mu}$ should
include not only the gravitational redshifts terms
$Z^{\lambda}_{\phantom{a}\mu}$ but also inertial coordinate twist
terms (ICT) $\tilde{Z}^{\alpha}_{\phantom{a}\lambda}$. Here the
first introduced ICT terms are different from the stretch or crush
effects of redshifts terms, and can be understood as wrinkle effects
on the free falling coordinates due to the inhomogeneous property of
gravitational field. And they can only make sense in gravitational
field and should be trivial(
$\tilde{Z}^{\beta}_{\phantom{a}\lambda}={\delta}^{\beta}_{\phantom{a}\lambda}$)
in flat spacetime. Generally speaking, $X^{\beta}_{\phantom{a}\mu}$
can be written as
$\tilde{Z}^{\beta}_{\phantom{a}\lambda}Z^{\lambda}_{\phantom{a}\mu}$.

 Because $\theta^{\alpha}_{\phantom{a}\mu}(x)$ denotes coordinate
transformations between LICS and the laboratory frame, we here give
a requirement:
\begin{equation}\label{integral}
\left\{ \begin{array}{ccc} \theta^{\alpha}_{\phantom{a}\mu,\nu} & = &  \theta^{\alpha}_{\phantom{a}\nu,\mu} \\
e^{\mu}_{\phantom{a}\alpha,\beta} & = &
e^{\mu}_{\phantom{a}\beta,\alpha}  \end{array} \right.
\end{equation}
It is quite natural to require that the coordinate transformations
are smooth and continuous, which can also be considered as a basic
assumption about nature. Now using eq.~(\ref{integral}) we can
derive the relationship between
$\Lambda^{\alpha}_{\phantom{a}\beta,\lambda}(x)$ and
$X^{\alpha}_{\phantom{a}\mu,\lambda}(x)$:
\begin{eqnarray}\label{condition}
\Lambda^{\alpha}_{\phantom{a}\beta,\nu}=\frac{1}{2}\Lambda^{\alpha}_{\phantom{a}\gamma}\eta^{\gamma\delta}(X^{-1\mu}_{\phantom{-1a}\beta}T_{\delta\nu\mu}
+X^{-1\lambda}_{\phantom{-1a}\delta}T_{\beta\lambda\nu}
-X^{-1\mu}_{\phantom{-1a}\beta}X^{-1\lambda}_{\phantom{-1a}\delta}
X^{\sigma}_{\phantom{a}\nu}T_{\sigma\mu\lambda})
\end{eqnarray}
in which $T_{\alpha,\nu,\mu}\equiv
X_{\alpha\nu,\mu}-X_{\alpha\mu,\nu}$. To look into the interesting
physics of the relatively complicated relation~(\ref{condition})
more clearly, we can choose some special coordinate conditions,
i.e.,
$X^{\alpha}_{\phantom{a}\mu}=\delta^{\alpha}_{\phantom{a}\mu}$,
eq.~(\ref{condition}) becomes:
\begin{equation}\label{simple1}
\Lambda^{\alpha}_{\phantom{a}\beta,\nu}=\Lambda^{\alpha}_{\phantom{a}\gamma}\eta^{\gamma\delta}
(X_{\delta\nu,\beta}-X_{\beta\nu,\delta})
\end{equation}
Continue requiring
$\Lambda^{\alpha}_{\phantom{a}\gamma}=\delta^{\alpha}_{\phantom{a}\gamma}$,
we finally get:
\begin{equation}\label{simple2}
\Lambda_{\alpha\beta,\nu}= X_{\alpha\nu,\beta}-X_{\beta\nu,\alpha}
\end{equation}
We can see that $\Lambda^{\alpha}_{\phantom{a}\beta,\nu}$ is totally
determined by $X^{\alpha}_{\phantom{a}\mu,\nu}$. And if we choose
$\alpha=0,\beta=1,\nu=0$, in the static Gravitational field, the
l.h.s. of eq.~(\ref{simple2}) is just acceleration, while the r.h.s.
is the derivative of gravitational potential with respect to $x^1$.
That is to say, the acceleration of a free falling particle in
static gravitational field at the instant of stationary is
determined by the gratitude of the gravitational potential.

The compatibility condition eq.~(\ref{condition}) gives no
constraint on $X_{\mu\nu}$. In order to get back to physical results
in flat spacetime, so here, before other compatibility conditions
are introduced, $\tilde{Z}^\beta_{\lambda}$ can be identified as
$\delta^\beta_{\lambda}$ and $\tilde{Z}^\beta_{\lambda,\rho}$ can be
set to zero. That is to say, $X_{\mu\nu}$
($X_{\mu\nu,\lambda}$)$|_{\tilde{Z}=\delta}$ equals to $Z_{\mu\nu}$
($Z_{\mu\nu,\lambda}$). Having got the
relationship~(\ref{condition}) between the derivatives of
$\Lambda^{\alpha}_{\phantom{a}\beta}$'s and
 $Z^{\alpha}_{\phantom{a}\mu}$'s, we can prove that the affine connection
$\Gamma^{\lambda}_{\mu\nu}=e^{\lambda}_{\phantom{a}\alpha}\theta^{\alpha}_{\phantom{a}\mu,\nu}$
takes the value of Christoffel Symbol in the free falling equation
\begin{equation}\label{gamma}
\frac{dU^{\lambda}}{d\tau}+
e^{\lambda}_{\phantom{a}\alpha}\theta^{\alpha}_{\phantom{a}\mu,\nu}U^{\mu
\nu}=0
\end{equation}

The proof is quite easy. Substituting the derivatives of
$\Lambda^{\alpha}_{\phantom{a}\beta}$ derived in
eq.~(\ref{condition}) into the definition of
$\Gamma^{\lambda}_{\mu\nu}$, then using the relation
$g_{\mu\nu}\equiv
Z^{\alpha}_{\phantom{a}\mu}\eta_{\alpha\beta}Z^{\beta}_{\phantom{a}\nu}$,
we finally get the results:
\begin{eqnarray*}
&&\Gamma^{\lambda}_{\mu\nu}|_{\tilde{Z}=\delta}\\
&=&e^{\lambda}_{\phantom{a}\alpha}\theta^{\alpha}_{\phantom{a}\mu,\nu}|_{\tilde{Z}=\delta}\\
&=&Z^{-1\lambda}_{\phantom{-1a}\gamma}\Lambda^{\phantom{a}\gamma}_{\alpha}(\Lambda^{\alpha}_{\phantom{a}\beta}Z^{\beta}_{\phantom{a}\mu})_{,\nu}\\
&=&Z^{-1\lambda}_{\phantom{-1a}\beta}Z^{\beta}_{\phantom{a}\mu,\nu}+Z^{-1\lambda}_{\phantom{-1a}\gamma}\Lambda^{\phantom{a}\gamma}_{\alpha}
\Lambda^{\alpha}_{\phantom{a}\beta,\nu}Z^{\beta}_{\phantom{a}\mu}\\
&=&Z^{-1\lambda}_{\phantom{-1a}\beta}Z^{\beta}_{\phantom{a}\mu,\nu}+\frac{1}{2}Z^{-1\lambda}_{\phantom{-1a}\gamma}\eta^{\delta\gamma}
(Z^{-1\omega}_{\phantom{-1a}\beta}T_{\delta\nu\omega}
+Z^{-1\xi}_{\phantom{-1a}\delta}T_{\beta\xi\nu}
-Z^{-1\omega}_{\phantom{-1a}\beta}Z^{-1\xi}_{\phantom{-1a}\delta}Z^{\sigma}_{\phantom{a}\nu}T_{\sigma\omega\xi})Z^{\beta}_{\phantom{a}\mu}\\
&=&Z^{-1\lambda}_{\phantom{-1a}\beta}Z^{\beta}_{\phantom{a}\mu,\nu}+\frac{1}{2}Z^{-1\lambda\delta}Z_{\delta\nu,\mu}
-\frac{1}{2}Z^{-1\lambda\delta}Z_{\delta\mu,\nu}
+\frac{1}{2}Z^{-1\lambda\delta}Z^{-1\xi}_{\phantom{-1a}\delta}Z^{\beta}_{\phantom{a}\mu}Z_{\beta\xi,\nu}
-\frac{1}{2}Z^{-1\lambda\delta}Z^{\beta}_{\phantom{a}\mu}Z_{\beta\nu,\delta}\\
&&-\frac{1}{2}Z^{-1\lambda\delta}Z^{\sigma}_{\phantom{a}\nu}Z_{\sigma\mu,\delta}
+\frac{1}{2}Z^{-1\lambda\delta}Z^{-1\xi}_{\phantom{-1a}\delta}Z^{\beta}_{\phantom{a}\nu}Z_{\beta\xi,\mu}\\
&=&\frac{1}{2}g^{\lambda\xi}[(Z^{\alpha}_{\phantom{a}\xi}\eta_{\alpha\beta}Z^{\beta}_{\phantom{a}\mu,\nu}
+Z^{\beta}_{\phantom{a}\mu}\eta_{\alpha\beta}Z^{\alpha}_{\phantom{a}\xi,\nu})
+(Z^{\alpha}_{\phantom{a}\xi}\eta_{\alpha\beta}Z^{\beta}_{\phantom{a}\nu,\mu}
+Z^{\alpha}_{\phantom{a}\xi,\mu}\eta_{\alpha\beta}Z^{\beta}_{\phantom{a}\nu})
-(Z^{\alpha}_{\phantom{a}\mu}\eta_{\alpha\beta}Z^{\beta}_{\phantom{a}\nu,\xi}
+Z^{\beta}_{\phantom{a}\nu}\eta_{\alpha\beta}Z^{\alpha}_{\phantom{a}\mu,\xi})]\\
&=&\frac{1}{2}g^{\lambda\xi}(g_{\xi\mu,\nu}+g_{\xi\nu\mu}-g_{\mu\nu,\xi})\\
&=&\left\{ \begin{array}{cc} \lambda & \\
 \mu & \nu    \end{array}   \right\}
\end{eqnarray*}
In the above proof we used eq.~(\ref{condition}) and the definition
$g_{\mu\nu}\equiv
Z^{\alpha}_{\phantom{a}\mu}\eta_{\alpha\beta}Z^{\beta}_{\phantom{a}\nu}$.
After the above relative lengthly discussion, we derived the same
free falling equations as those in General Relativity.

\section{Differential Equations}

In this section we will demonstrate, according to the compatibility
conditions of coordinates, $\tilde{Z}_{\beta\lambda}$ and any order
of its derivatives can be calculated exactly, provided that
$Z_{\lambda\mu}$ and its any order of derivatives are known.
Differentiating with respect to $\mu$ on both sides of
eq.~(\ref{condition}) and requiring
$\Lambda^{\alpha}_{\phantom{a}\beta,\mu,\nu}=\Lambda^{\alpha}_{\phantom{a}\beta,\nu,\mu}$,
we get the following equation:

\begin{eqnarray}\label{con}
&&[Z^{-1\rho}_{\phantom{-1\rho}\beta}(Z^{\gamma}_{\phantom{\gamma}\mu}\tilde{Z}_{\alpha\gamma,\rho,\nu}
-Z^{\gamma}_{\phantom{\gamma}\nu}\tilde{Z}_{\alpha\gamma,\rho,\mu})
+Z^{-1\rho}_{\phantom{-1\rho}\beta}(Z^{\sigma}_{\phantom{\gamma}\mu}\tilde{Z}_{\sigma\alpha,\rho,\nu}
-Z^{\sigma}_{\phantom{\gamma}\nu}\tilde{Z}_{\sigma\alpha,\rho,\mu})\nonumber\\
&+&Z^{-1\lambda}_{\phantom{-1\rho}\alpha}(Z^{\gamma}_{\phantom{\gamma}\nu}\tilde{Z}_{\beta\gamma,\lambda,\mu}
-Z^{\gamma}_{\phantom{\gamma}\mu}\tilde{Z}_{\beta\gamma,\lambda,\nu})
+Z^{-1\lambda}_{\phantom{-1\rho}\alpha}(Z^{\sigma}_{\phantom{\gamma}\nu}\tilde{Z}_{\sigma\beta,\lambda,\mu}
-Z^{\sigma}_{\phantom{\gamma}\mu}\tilde{Z}_{\sigma\beta,\lambda,\nu})]-W_{\alpha\beta\mu\nu}=0.
\end{eqnarray}

In the above equation we introduced a tensor variable
$W_{\alpha\beta\mu\nu}$ for convenience. And it is defined as:
\begin{eqnarray}
W_{\alpha\beta\mu\nu}&=&
[Z^{-1\rho}_{\phantom{-1\rho}\beta}(Z_{\alpha\nu,\rho,\mu}-Z_{\alpha\mu,\rho,\nu})
+Z^{-1\lambda}_{\phantom{-1\rho}\alpha}(Z_{\beta\mu,\lambda,\nu}-Z_{\beta\nu,\lambda,\mu})\nonumber\\
&+&Z^{-1\rho}_{\phantom{-1\rho}\beta}Z^{-1\lambda}_{\phantom{-1\rho}\alpha}Z^{\sigma}_{\phantom{\sigma}\mu}
(Z_{\sigma\rho,\lambda,\nu}-Z_{\sigma\lambda,\rho,\nu})
+Z^{-1\rho}_{\phantom{-1\rho}\beta}Z^{-1\lambda}_{\phantom{-1\rho}\alpha}Z^{\sigma}_{\phantom{\sigma}\nu}
(Z_{\sigma\lambda,\rho,\mu}-Z_{\sigma\rho,\lambda,\mu})]\\
&+&[2\Lambda_{\sigma\alpha}\Lambda^{\sigma\delta}_{\phantom{\sigma\delta},\mu}
\Lambda_{\eta\delta}\Lambda^{\eta}_{\phantom{\delta}\beta,\nu}+Z^{-1\rho}_{\phantom{-1\rho}\beta,\mu}T_{\alpha\nu\rho}
+Z^{-1\lambda}_{\phantom{-1\rho}\alpha,\mu}T_{\beta\lambda\nu}
-(Z^{-1\rho}_{\phantom{-1\rho}\beta}Z^{-1\lambda}_{\phantom{-1\rho}\alpha}Z^{\sigma}_{\phantom{\sigma}\nu})_{,\mu}
T_{\sigma\rho\lambda}-(\mu\leftrightarrow\nu)]\nonumber
\end{eqnarray}

It is clear that eq.(\ref{con}) is a linear equation of
$\tilde{Z}_{\mu\nu,\lambda,\rho}$. So we can rewrite the above
equation as the form:
\begin{equation}\label{con2}
\tilde{Z}_{\mu\nu,\lambda,\rho}=L_{2}(\tilde{Z}_{\mu\nu,\lambda,\rho}),
\end{equation}

in which the function $L_{2}$ is some kind of linear function.
Similar to what we did to eq.~(\ref{condition}) and remembering
$\tilde{Z}_{\mu\nu,\lambda,\rho,\delta}=\tilde{Z}_{\mu\nu,\lambda,\delta,\rho}$,
we get the 3rd order  compatibility condition

\begin{equation}
\tilde{Z}_{\mu\nu,\lambda,\rho,\delta}=L_{3}(\tilde{Z}_{\mu\nu,\lambda,\rho,\delta})
\end{equation}

Using this iterative method we can get the compatibility conditions
of $\tilde{Z}_{\mu\nu}$'s any order derivatives.
\begin{equation}\label{conn}
\tilde{Z}_{\mu\nu,\lambda_{1}...\lambda_{n}}=L_{n}(\tilde{Z}_{\mu\nu,\lambda_{1}...\lambda_{n}}).
\end{equation}

We may decompose eq.(\ref{conn}) as follows :
\begin{equation}\label{nconn}
\tilde{Z}_{\mu\nu,\lambda_{1}...\lambda_{n}}=\bar{L}_{n}(\tilde{Z}_{\mu\nu,\lambda_{1}...\lambda_{n}})
+Y_n(Z_{\mu\nu},\cdots,Z_{\mu\nu,\lambda_1...\lambda_n}),
\end{equation}
in which $\bar{L}_n$ is the homogeneous term and $Y_n$ is the
nonhomogeneous term. When $Y_n$ takes all possible values, the
solutions of eq.(\ref{nconn})
$\{\tilde{Z}_{\mu\nu,\lambda_{1}...\lambda_{n}}\}$ form a group $G$
under the action of tensor addition. While the solutions of the
corresponding homogeneous linear equation
\begin{equation}\label{lcon}
 \tilde{Z}_{\mu\nu,\lambda_{1}...\lambda_{n}}=\bar{L}_{n}(\tilde{Z}_{\mu\nu,\lambda_{1}...\lambda_{n}})
\end{equation}
form a subgroup of $G$.

Before solving the above equations, we first discuss our definition
of spacetime curvature. As we all know, in General Relativity, the
curvature of spacetime is described by $R_{\kappa\omega\mu\nu}$. But
in our calculations, the definition is not sufficient. The term
$\tilde{Z}_{\mu\nu}$ appears when there is spacetime curvature and
vanishes in flat spacetime. According to eq.(\ref{nconn}), a nonzero
$\{Y_n\}$ requires nontrivial $\tilde{Z}_{\mu\nu}$. If $\{Y_n\}$ is
zero, Minkowski coordinates is included in this case,
$\tilde{Z}_{\mu\nu}$ must equal to $\eta_{\mu\nu}$ to recover the
Minkowski spacetime. Hence, at one spacetime point
$R_{\kappa\omega\mu\nu}=0$ is not an accurate definition of
flatness, and it should be defined as $\{Y_n\}=0$. Meanwhile at rank
2 $Y_2$ is just $W_{\alpha\beta\mu\nu}$, which is equivalent to
curvature tensor $R_{\kappa\omega\mu\nu}$(see eq.(\ref{wrelation})).

To find the physical solution of the above linear tensor equation
eq.(\ref{nconn}), we first consider the solution of eq.(\ref{lcon})
in which the absence of nonhomogeneous term means that the space
time is flat. For the ICT term only appears when there is gravity
and it is trivial in flat spacetime, it means that only one trivial
solution of eq.(\ref{lcon}), $\tilde{Z}_{\mu\nu}=\eta_{\mu\nu}$, is
physical. The properties of a group and its subgroup ensure that the
physical solution of eq.(\ref{nconn}) is unique.

We here in this paper give the solution of $L_2$ eq.(\ref{con2}).
First we derived some properties of the newly defined tensor
variable $W_{\alpha\beta\mu\nu}$ :
\begin{eqnarray}\label{wrelation}
W_{\alpha\beta\mu\nu} &=& -W_{\beta\alpha\mu\nu}\nonumber\\
W_{\alpha\beta\mu\nu} &=& -W_{\alpha\beta\nu\mu}\nonumber\\
W_{\alpha\beta\mu\nu} &=&
Z^{\gamma}_{\phantom{a}\mu}Z^{\sigma}_{\phantom{a}\nu}Z^{-1\rho}_{\phantom{-1a}\alpha}
Z^{-1\lambda}_{\phantom{-1a}\beta}W_{\gamma\sigma\rho\lambda}\nonumber\\
W_{\alpha\beta\mu\nu} &=&
\frac{1}{2}\{[Z^{\gamma}_{\phantom{a}\mu}Z^{-1\rho}_{\phantom{-1a}\alpha}W_{\gamma\beta\rho\nu}
-(\alpha\leftrightarrow\beta)]-(\mu\leftrightarrow\nu)\}\nonumber\\
W_{\alpha\beta\mu\nu} &=&
-2Z^{-1\kappa}_{\phantom{-1a}\alpha}Z^{-1\omega}_{\phantom{-1a}\beta}R_{\kappa\omega\mu\nu}
\end{eqnarray}
Here the term $R_{\kappa\omega\mu\nu}$ is the Riemann curvature.
Using the above relation of $W_{\alpha\beta\mu\nu}$, it is easy to
check that the solution of eq.(\ref{con}) have the form:
\begin{equation}\label{ztilde2}
\tilde{Z}_{\alpha\mu,\beta,\nu}=\frac{1}{12}(Z^{\rho}_{\phantom{a}\beta}
Z^{-1\lambda}_{\phantom{-1a}\mu}
W_{\alpha\rho\lambda\nu}+Z^{\rho}_{\phantom{a}\nu}
Z^{-1\lambda}_{\phantom{-1a}\mu} W_{\alpha\rho\lambda\beta})
\end{equation}

As all the variables describing coordinate transformations is clear,
we can build the coordinate transformation rules in general. The
definition of covariant derivative on cotracovariant tensor is:
\begin{eqnarray}
&& A_{\alpha_1\alpha_2\cdots\alpha_n,\beta}\nonumber\\
 &=& (\frac{\partial x^{\mu_1}}{\partial\xi^{\alpha_1}}\cdots
\frac{\partial x^{\mu_n}}{\partial\xi^{\alpha_n}}A_{\mu_1\cdots\mu_n})_{,\nu}\frac{\partial x^{\nu}}{\partial\xi^{\beta}}\nonumber\\
&=& e^\nu_{\phantom{a}\beta} \prod_j e^{\mu_j}_{\phantom{a}\alpha_j}
( A_{\mu_1\cdots\mu_n ,\nu} - \sum^n_{i=1}\Gamma^{\rho_i}_{\mu_i
\nu}
A_{\mu_1\cdots\rho_i\cdots \mu_n})\nonumber\\
&=& e^\nu_{\phantom{a}\beta} \prod_j e^{\mu_j}_{\phantom{a}\alpha_j}
A_{\mu_1\cdots\mu_n ;\nu}
\end{eqnarray}

in which $\Gamma^{\rho_i}_{\mu_i
\nu}=e^{\rho_i}_{\phantom{a}\sigma}\theta^\sigma_{\phantom{a}\mu_i
,\nu}$ is defined in  the free falling equation eq.(\ref{gamma}).

For the covariant tensor, a similar calculation can be performed and
we come to a similar result:
\begin{eqnarray}
&& A^{\alpha_1\alpha_2\cdots\alpha_n}_{\phantom{aaaaaaaa},\beta}\nonumber\\
 &=& (\frac{\partial \xi^{\alpha_1}}{\partial x^{\mu_1}}\cdots
\frac{\partial \xi^{\alpha_n}}{\partial x^{\mu_n}}A^{\mu_1\cdots\mu_n})_{,\nu}\frac{\partial x^{\nu}}{\partial\xi^{\beta}}\nonumber\\
&=& e^\nu_{\phantom{a}\beta} \prod_j
\theta^{\alpha_j}_{\phantom{a}\mu_j} (
A^{\mu_1\cdots\mu_n}_{\phantom{aaaaaa},\nu} +
\sum^n_{i=1}\Gamma^{\mu_i}_{\rho_i \nu}
A^{\mu_1\cdots\rho_i\cdots \mu_n})\nonumber\\
&=& e^\nu_{\phantom{a}\beta} \prod_j
\theta^{\alpha_j}_{\phantom{a}\mu_j}
A^{\mu_1\cdots\mu_n}_{\phantom{aaaaaa};\nu}
\end{eqnarray}

As we know in General Relativity, the second order covariant
differentiation of a vector or tensor field is not symmetric about
the differential indices, instead there is an extra term which
couples to spacetime curvature,

\begin{equation}
A_{\mu;\nu;\lambda}=A_{\mu;\lambda;\nu}-A_{\sigma}
R^{\sigma}_{\phantom{\sigma} \mu \nu \lambda}.
\end{equation}

But in our case, the term $A_{\mu;\nu;\lambda}$ is just the
coordinate transformed form of the second order partial derivative
term in local Lorentz coordinate $A_{\alpha,\beta,\gamma}$, and it
is certainly symmetric about the factor $\nu$ and $\lambda$.
\begin{eqnarray}
A_{\mu;\nu;\lambda}&=&\frac{\partial\xi^{\alpha}}{\partial x^{\mu}}
\frac{\partial\xi^{\beta}}{\partial x^{\nu}}
\frac{\partial\xi^{\gamma}}{\partial
x^{\lambda}}A_{\alpha,\beta,\gamma}\\
&=&A_{\mu,\nu,\lambda}-\Gamma^{\xi}_{\mu\nu,\lambda}A_{\xi}-\Gamma^{\xi}_{\mu\nu}A_{\xi,\lambda}
-\Gamma^{\xi}_{\mu\lambda}A_{\xi,\nu}+\Gamma^{\eta}_{\mu\lambda}\Gamma^{\xi}_{\eta\nu}A_{\xi}
-\Gamma^{\xi}_{\nu\lambda}(A_{\mu;\xi})
\end{eqnarray}
In the above equation we get a very close form as that in General
Relativity. The only difference lies in the term
$\Gamma^{\xi}_{\mu\nu,\lambda}$. In our cases, it must be
 treated very carefully due to the contribution of the
 second order differentiation of $\tilde{Z}$.
\begin{eqnarray}
\theta^{\alpha}_{\phantom{a}\mu,\nu,\lambda}
&=&(\Lambda^{\alpha}_{\phantom{a}\beta} X^{\beta}_{\phantom{a}\mu})_{,\nu,\lambda}\nonumber\\
&=&\Lambda^{\alpha}_{\phantom{a}\beta,\nu,\lambda}X^{\beta}_{\phantom{a}\mu}
+\Lambda^{\alpha}_{\phantom{a}\beta}(\tilde{Z}^{\beta}_{\phantom{a}\rho}Z^{\rho}_{\phantom{a}\mu})_{,\nu,\lambda}
+\Lambda^{\alpha}_{\phantom{a}\beta,\nu}X^{\beta}_{\phantom{a}\mu,\lambda}
++\Lambda^{\alpha}_{\phantom{a}\beta,\lambda}X^{\beta}_{\phantom{a}\mu,\nu}
\end{eqnarray}

\section{Conclusion}
As a general concept, this paper presents a description of LICS,
which can be understood as a new interpretation of Equivalence
Principle. Having got the right description of LICS, all the
calculations concerning gravitational effects are based on
coordinate transformations which are required smooth and continuous.
The form of $\Gamma^{\lambda}_{\mu \nu}$ can be exactly calculated
once the redshifts of spacetime $Z_{\mu\nu}$ is known, but it is not
Christoffel symbol as that in General Relativity. Then we derived
the same free falling equations as those in General Relativity.
However, the definitions of second and higher order covariant
derivatives are different from those in General Relativity. Under
this interpretation, many concepts of gravitation may be changed
including the gravitational field equations. We shall focus on this
point in our forthcoming coming paper\cite{zx}. In conclusion, our
main point in this paper is to give a description of LICS, and all
the gravitational effects can be calculated through this coordinate
transformation.

%\newpage %Just because of unusual number of tables stacked at end
\bibliographystyle{hunsrt}
%\bibliography{zxep}% Produces the bibliography via BibTeX.

\end{document}